\documentstyle[twocolumn,prl,aps,epsf,subeqn,cite]{revtex}
\def\be{\begin{equation}}
\def\ee{\end{equation}}
                              \def\barr{\begin{array}}
                              \def\earr{\end{array}}

\def\etal{{\em et al.}}

                              \def\gev{\: {\rm GeV} }
                              \def\tev{\: {\rm TeV} }
                              
                              \def\fb{\: {\rm fb}}
\def\gappeq{\mathrel{\rlap {\raise.5ex\hbox{$>$}}
            {\lower.5ex\hbox{$\sim$}}}}
\def\lappeq{\mathrel{\rlap{\raise.5ex\hbox{$<$}}
            {\lower.5ex\hbox{$\sim$}}}}
\def\ra{\rightarrow}

\begin{document}                                                              


\title{Higgs Boson Signals in Three $b$-jet Final States at the
Fermilab Tevatron}

\author{Debajyoti Choudhury$^{1)}$, Anindya Datta$^{2)}$ and
	Sreerup Raychaudhuri$^{3)}$}
\address{$^{1)}$Mehta Research Institute, 
Chhatnag Road, Jhusi, Allahabad 211 019, India.
Electronic address:debchou@mail.cern.ch.\\
$^{2)}$Department of Physics, University of Calcutta, 92
A.P.C.  Road, Calcutta 700 009, India.\\
$^{3)}$Theory Division, CERN, CH 1211 Geneva 23, Switzerland.
Electronic address:sreerup@mail.cern.ch.}

\maketitle
\begin{abstract}
\hspace*{-0.65cm}
At the Fermilab Tevatron, final states with three tagged $b$-jets
could play an important role in searches for a Higgs boson with mass
in the range 100--300 GeV. These signals arise from $gb$ fusion and
we demonstrate their observability in the limit of a large $b$-quark
Yukawa coupling. Rather promising discovery limits on such a coupling
are obtained and consequent effects on the parameter space of the
Higgs-boson sector in the MSSM are discussed.
\\[0.1cm]
PACS number(s): 12.60.-i, 14.80.Cp, 14.65.Fy, 12.38.Bx
\end{abstract}
\hspace*{0.65cm}


\begin{narrowtext}

The Higgs boson is the last ingredient of the Standard Model (SM)
that still awaits discovery. It is also a crucial component since the
mechanism for electroweak symmetry-breaking --- and hence the masses
of weak gauge bosons and fermions --- arises from its interactions.
There exist strong indications from the fitting of electroweak
parameters to available data that the Higgs boson has a relatively
small mass~\cite{ErlLan}, which makes it likely to be seen either at
the Fermilab Tevatron (in its future runs) or at the CERN LHC (when
it becomes operational). Curiously, although failure to find a
Higgs boson at all is likely to require serious rethinking of current
theories of electroweak interactions, establishing its existence
could also raise a rash of questions which might not be easy to answer.
For example, one could ask whether a discovered Higgs boson is the
one predicted in the SM or whether it is an ingredient of some
extended theory with extra symmetries. The easiest way to settle this
question would be by the detection of (an) additional Higgs boson(s)
such as is (are) predicted in several of these
models. Alternatively, on discovering a Higgs boson, 
we could determine its couplings and compare these 
with theoretical predictions within a given model.
With the higher integrated luminosity expected in
Run II of the Tevatron, the requisite precision in the 
measurements should be available,  at least for
favourable regions of the parameter space. A truly dramatic
way of discovering new physics beyond the SM would be the observation
of Higgs-boson signals where none are predicted in the SM.

Enhanced Yukawa couplings for the $b$-quark in some models beyond the
SM open up many such interesting new channels. At the CERN LEP, for
example, $e^+ e^- \ra Z^* \ra b \bar b A$
could help\cite{WelKan} extend the Higgs-boson
discovery limits beyond the expectation from the study of
Higgs-strahlung processes, which are the main avenue of Higgs-boson
searches at present. At the Fermilab Tevatron, similar processes have
been studied, some recent efforts being those of
Refs.~\cite{DreGucRoy,DiaHeTaiYua,CarMreWag}. The
consensus appears to be that detection of the SM Higgs boson radiated
off a heavy quark would be difficult, but (non-SM) Higgs bosons with
enhanced Yukawa couplings would be much better candidates for a study
in $4b$ and $bb\tau\tau$ final states.

In this letter, we propose the related process
\be
   g b(\bar b) \ra \phi b(\bar b)  
                                      \label{eq:process}
\ee
as a possible production channel for a generic Higgs boson,
$\phi$. While the analogous interaction for the charged
Higgs boson (in two Higgs-doublet models) has been studied
extensively in the context of the CERN LHC, this particular process
has been neglected so far, probably under the impression that the
cross section would be hopelessly small. This is, in fact, true for
the SM Higgs boson, unless we consider the TeV-33 option 
at the Fermilab Tevatron. 
However, we shall show that it need not be the
case -- even in Run II  -- for theories
with an enhanced $\phi b \bar b$ coupling.

The Yukawa interaction can be written ${\cal L}_{\phi b \bar b} = h_b
\phi \bar b b$, or ${\cal L}_{\phi b \bar b} = ih_b \phi \bar b
\gamma_5 b$, depending on the $CP$ assignments of the scalar $\phi$.
Given such an interaction, the cross-sections (to leading order)
for the processes in Eq.~(\ref{eq:process}) scale as the
ratio $(h_b/h_b^{\rm SM})^2$, irrespective of the $CP$ properties of the
field $\phi$. The precise value of the ratio $h_b/h_b^{\rm SM}$
depends, of course, on the model. In the minimal supersymmetric
extension of the SM (MSSM), for example, $h_b/h_b^{\rm SM} =
\tan\beta, -\sin\alpha /\cos\beta$ and $\cos\alpha /\cos\beta$ for
the pseudoscalar ($A^0$), light scalar ($h^0$) and heavy scalar
($H^0$) Higgs bosons respectively, where the angles $\alpha,\beta$
have their usual meanings\cite{DawGunHabKan}. In the large $\tan
\beta$ limit, the coupling to the $A^0$ (obviously) and the $H^0$
are strongly enhanced; that to the $h^0$ undergoes a more
modest increase. Similar enhancements of the $b$-quark Yukawa
couplings can occur in composite models such as those with
topcolour-assisted technicolour~\cite{topcolor}.

Once a $\phi$ has been produced, various decay channels are available
to it. Again, there is considerable model dependence. For the SM
$H^0$, the major two-body decay modes are $H^0 \ra b \bar b, c \bar
c, \tau^+ \tau^-$ with relative branching ratios of $23:2:4$
approximately\cite{comment:branching}. While these modes dominate for
low Higgs-boson masses, the $H^0 \ra W W^\ast$ mode becomes
increasingly important for $m_H > 100 \gev$. In the MSSM, on the
other hand, the $WW^\ast$ (or $Z Z^\ast$) mode is inaccessible to the
$A^0$. Consequently, $Br (A^0 \ra b \bar b) \gappeq 90 \%$ (unless
$A^0$ can decay into a pair of superpartners). Similarly, in the
event of $\tan\beta \gg 1$, there are just three important decay
modes for the scalar $h^0$, namely $h^0 \ra b \bar b, \tau^+ \tau^-,
W W^\ast$, the first always accounting for more than 70\%. While
the $W W^\ast$ mode may also be profitably used as a probe of
Higgs-boson interactions, we prefer to concentrate on the simpler
two-body decays of the Higgs boson into fermions. Thus -- for the SM
and MSSM at least -- the final states of greatest interest in $gb$
fusion are $b \tau^+ \tau^-$ and $b b \bar b$~\cite{comment:gam_gam}.
Of these two, the former has smaller backgrounds, whether 
from QCD or from associated $Z$-production.
However, $Br (\phi \ra \tau^+ \tau^-)$ is 
suppressed approximately by a factor
of $6$ or more over most of the parameter space. 
More importantly, invariant mass reconstruction for a
$\tau$-pair is difficult because of multiple neutrinos
arising in tau-pair decays. As mass reconstruction turns out to be a major
tool in the isolation of a Higgs-boson signal, we shall not comment
on the $\tau^+ \tau^- $ channel any further\cite{comment:tautau}.

Thus, our choice of the final state is  ($b b \bar b$) or ($b \bar b
\bar b$), which we generically denote by $3b$. The signal cross
section may be written as $\sigma(gb \ra 3b) = {\cal R}^2 \sigma_{\rm
SM}(gb \ra 3b)$, where
\be
{\cal R} = \left( \frac{h_b}{h_b^{\rm SM}} \right)
          \; \left[ \frac{Br(\phi^0 \;\ra \; b \: \bar{b})}
                {Br(H^0_{\rm SM} \;\ra \; b \:\bar{b})}
             \right]^{1/2} \ .
\ee
The advantage of using ${\cal R}$ as a free parameter is obvious -- it
contains the entire model-dependence of the cross section and hence
enables us to make a {\em model-independent} study of the $3b$
signal. 

At this point, it becomes necessary to comment 
on the size of $h_b^{\rm SM}$.  The 
low-energy value can be inferred from the pole mass,
for which we use $m_b = 4.3~\gev$~\cite{PDG}. At large momentum
transfers, QCD corrections can be important. Since the complete set
of corrections have not been calculated, we include only the
suppression due to the running of $m_b$. As additional QCD
corrections usually tend to increase the cross-section, our results
should be regarded as a {\it conservative} estimate of the signal.
Our expressions are consistent with those in Ref.\cite{HDECAY}.

The backgrounds to (\ref{eq:process}) arise from two main sources: ($i$)
`authentic' $3b$ events from QCD and/or weak interactions; ($ii$)
spurious events of the type  $2b+J, b+2J$ or $3J$,
where $J$ denotes a non-$b$ jet {\em misidentified} as a $b$ jet.
Herein lies the advantage of considering a $3b$ signal:
misidentification probabilities are usually low and `authentic' $3b$
backgrounds carry the same suppression from the $b$-quark flux as the
signal. In contrast to this, $4b$ backgrounds~\cite{DiaHeTaiYua}
 could be generated by QCD processes
arising from valence quarks or gluons, which have enormous fluxes by
comparison.

The large number of diagrams contributing to the background are
calculated using the helicity amplitude package {\sc
madgraph}~\cite{MadGraph}. 
To estimate the number of events and their distribution(s), we use 
a parton-level Monte-Carlo event generator. For the parton 
densities in the proton, we use the {\sc cteq3-m} structure
functions as incorporated in the package {\sc pdflib}~\cite{Pdflib}.

Since the QCD background is populated mostly at low transverse
momenta ($p_T$) and high rapidity ($\eta$) of the jets, we demand
that the final state be composed of {\em exactly} three hard jets
($j$) with $ p_T^j > 20 \gev, \; |\eta_j|  < 2$. 
We also require that the angular separation of the jets be
substantial, {\em i.e.} 
$ \Delta R_{jj} \equiv 
  \sqrt{ (\Delta \eta_{jj})^2 + (\Delta \phi_{jj})^2 } > 1.0  \ , $ 
adapting the well-known cone algorithm for jet separation to a
parton-level analysis. While $\Delta R_{jj} > 0.7$ is usually
considered sufficient for jets to be separable,
the more stringent cut (especially for the two
softer jets) eliminates a significant portion of the QCD background
without affecting the signal at all. To eliminate the background from
$gb \ra bZ^0$, we demand that 
all events where the invariant mass 
$m_{ij}$ satisfies $ 80 \gev <  m_{ij} < 100 \gev$, 
for {\em any} of the three possible pairings ($ij$), 
should be rejected from the analysis. Similarly, a requirement of
$ m_{ij} > 10 \gev $ helps to further reduce the QCD background.
Once these kinematic cuts are applied, we are in a position to
utilize a particular feature of the signal event topology. The
final-state $b$ in the production process $gb \ra b\phi$ tends to be
collinear with the initial $b$ quark and hence (usually) to have a
transverse momentum smaller than those of the $b$'s arising from
the scalar decay $\phi \ra b\bar b$. If we label the $b$-jets
according to their $p_T$, thus: $p_T(j_1) > p_T(j_2) >
p_T(j_3) $ --- the invariant mass $m_{12}$ of the pair of hardest jets
will reconstruct to the Higgs-boson mass for a majority of the signal
events. Of course, there will always be some events where the pair
(12) does not originate from the scalar, but such configurations are
subdominant and become progressively so for larger $m_\phi$. Thus the
$m_{12}$ distribution for the signal will exhibit a
characteristic peak, illustrated in Fig.~\ref{fig:invmass}. 
With the elimination of the $Z$ events, the
background does not show any such peaking.
 
\begin{figure}[htb]
\vspace*{-0.0cm}
\hspace*{-1.0cm}
\epsfxsize=9.3cm\epsfysize=5.3cm
\epsfbox{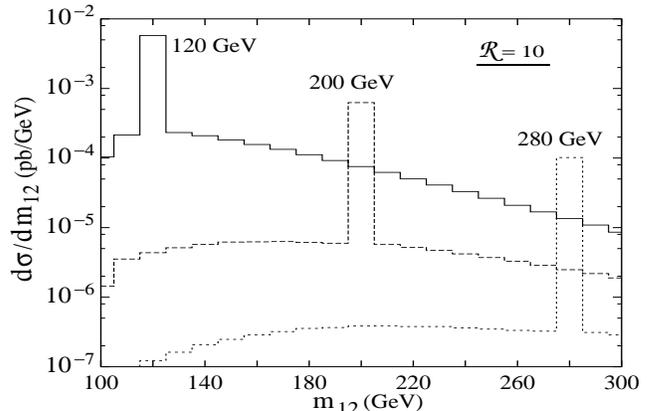}\\[-0.3cm]
\caption[fig:fig1]{Distribution in the invariant mass of the
pair of $b$-jets with the largest $p_T$. Solid, dashed and dotted
lines correspond to three different masses (marked) of the Higgs boson.
} \label{fig:invmass}
\end{figure}
 
We can thus use  $m_{12}$ as a discriminator for
Higgs-boson resonances. When looking for a $\phi$ with a given mass,
it is necessary to select only those events that lead to an $m_{12}$
close to $m_\phi$. Closeness is usually quantified
by the invariant mass resolution which is taken to be~\cite{Tev-2000}
${\rm max} \{ 10~\gev, \sqrt{0.8 \gev \times
m_\phi} \}$ for the entire range $m_\phi =$ 100--300 GeV.
We note
in passing that similar -- but much weaker  -- peaks can be seen in the
other two mass distributions $m_{13}, m_{23}$, although we have not
exhibited them.

Even with this kinematic selection procedure, the background is
orders of magnitude larger than the signal. We demand, therefore,
that {\em all} three $b$'s be tagged. Of course, progressively smaller
fractions of $2b+J,b+2J$ and $3J$ states, respectively,
will be misidentified as 
$3b$ states. To quantify this, for each individual jet, we 
use~\cite{Tev-2000} a $b$-tagging efficiency $\epsilon_b = 0.6$ and 
misidentification probability $P_{\rm mis} = 0.005$.
This eliminates most of the ($2b+J$) and almost all of the $b+2J$ and
$3J$ backgrounds while suppressing the signal (and the `authentic'
$3b$ backgrounds) by a factor of $(0.6)^3 \simeq 0.2$.

Using all these cuts and selection criteria, we are now in a position
to compare signal with background. Fig.~\ref{fig:eventnum} shows the
distribution in invariant mass $m_{12}$, assuming uniform bins of 10
GeV\cite{comment:resolution} in the entire range $m_{12} = 100$--$300
\gev$, for an integrated luminosity of $L = 1 \fb^{-1}$.   
To illustrate the usefulness of
this channel, we have considered a somewhat optimistic value ${\cal R}
= 50$, for which it is obvious that the signal is significantly
larger than the background. However, since it is only necessary for
the signal to exceed the {\em fluctuation} in the background at (say)
95\% confidence level (CL), the actual reach in ${\cal R}$
encompasses much lower values.

\begin{figure}[htb]
\vspace*{-0.0cm}
\hspace*{-1.0cm}
\epsfxsize=9.3cm\epsfysize=5.3cm
\epsfbox{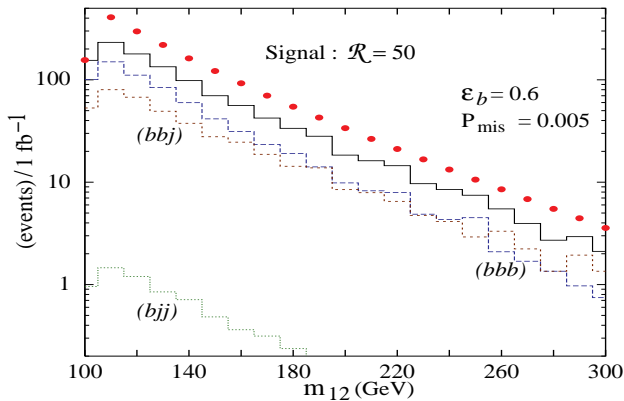}\\[-0.3cm]
\caption[fig:fig2]{The number of events in each $m_{12}$ bin for
signal (points) and the total background (solid line). The individual
contributions to the background are also shown (dashed and dotted lines).
} \label{fig:eventnum}
\end{figure}

Fig.~\ref{fig:eventnum} also shows the individual channels
contributing to the background. For relatively low Higgs-boson
masses, the largest background arises from `authentic' $3b$ processes,
single misidentified non-$b$ jets come in second and double
misidentification being strongly suppressed. Triple misidentification
is severely suppressed and need not concern us
further. For higher masses of the Higgs boson, however, the
single-misidentified-jet background catches up with the `authentic'
one, so that the importance of this channel hardly needs to be
emphasized.

We are now in a position to speculate on the mass reach of this mode.
Concentrating on the appropriate $m_{12}$ bin, we can evaluate the
minimum value of ${\cal R}$ that would allow us to establish/exclude
a Higgs boson with the corresponding mass. This is done in a
straightforward manner by comparing the signal size (in that bin)
with the statistical fluctuation in the background (in that bin),
calculated using Poisson statistics.

In Fig.~\ref{fig:exclusion}, we exhibit 95\% CL discovery limits for
three expected luminosities at the Fermilab Tevatron. If no excess in
$m_{12}$ is seen, the parameter space {\em above} the curves can be
ruled out at the appropriate confidence level. That the minimal value
of ${\cal R}$ grows with $m_\phi$ is expected, since larger $m_\phi$
leads to small cross sections for the process (\ref{eq:process}). The
minimum at $m_\phi \sim 110 \gev$ is an artefact of the kinematic cut
designed to eliminate the $Z$ background (in the
absence of such a cut, of course, all the discovery limits would be
much worse). It is possible, however,
that a more sophisticated analysis could improve the bounds for
$m_\phi \sim m_Z$. The curve for a luminosity of 100 pb$^{-1}$ 
represents the region which can be ruled out by existing data.

\begin{figure}[htb]
\vspace*{-0.0cm}
\hspace*{-1.0cm}
\epsfxsize=9.3cm\epsfysize=5.3cm
\epsfbox{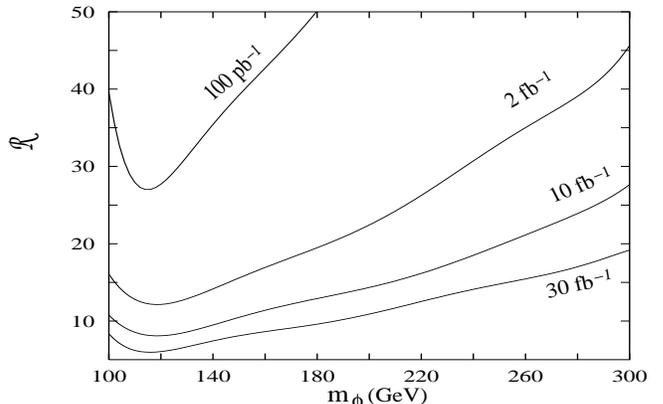}\\[-0.3cm]
\caption[fig:fig3]{Model-independent discovery limits 
for a given luminosity. The area above the curves can be
excluded at 95\% CL. }
\label{fig:exclusion}
\end{figure}

In obtaining Fig.~\ref{fig:exclusion}, we have assumed that only one
$\phi$ lies in the mass bin under consideration. If this is not 
the case, or if there is more than one Higgs-boson state in the region under 
consideration, then the constraints originating from all
such $\phi$'s will have to be compounded appropriately.
This could be parametrized as an enhancement in the
effective value of ${\cal R}$, with the actual increase 
determined by the model parameters. The curves in
Fig.~\ref{fig:exclusion} are thus conservative and, as we have
already stressed, {\em model-independent}.

In view of the intrinsic interest in the MSSM, we now pass from 
the general to the particular and derive explicit constraints on 
this scenario. Before presenting our results, some preliminary 
remarks are in order. Given the current bounds on sfermions, 
it is clear that it is not possible for a 100--300 GeV Higgs boson 
to decay into a pair of these. However, bounds on charginos and 
neutralinos are much weaker, and hence, we need to make some 
assumptions about the Wino mass parameter $M_2$ and the higgsino 
mixing parameter $\mu$. In our analysis we choose two illustrative 
sets of such parameters.

With the respective values of ${\cal R}$ 
determined by $m_A$, $\tan \beta$, $M_2$, $\mu$, the 
sfermion masses and the stop mixing parameter $\tilde A_t$~\cite{HDECAY},
the three individual constraints can now be combined.
The resultant bounds in the $m_A$--$\tan \beta$ 
plane are presented in Fig.~\ref{fig:matb}.
Solid curves correspond to the case where all the superpartners
are very heavy, thereby reducing the scenario to a constrained 
two-Higgs doublet model; dashed curves correspond to relatively light 
superpartners. With three non-trivial contributions (more than) offsetting 
any suppression due to branching ratios, the bounds on $\tan \beta$ 
are analogous to those of Fig.~\ref{fig:exclusion}.
Interestingly, the dependence on the MSSM parameters is rather
weak, as evidenced by the small difference in the two sets.
This is not unexpected as light Higgs bosons hardly decay into
the superpartners, even when the mass parameters are set as low as 
150 GeV. Radiative corrections in the Higgs sector, on the other hand, 
can significantly alter the individual couplings. On summing
over the three contributions, though,  the residual effects are small
and are not easily visible on the scale of Fig.~\ref{fig:matb}.
Our predictions are thus quite robust.
\begin{figure}[tb]
\vspace*{-0.2cm}
\hspace*{-1.0cm}
\epsfxsize=9.3cm\epsfysize=5.3cm
\epsfbox{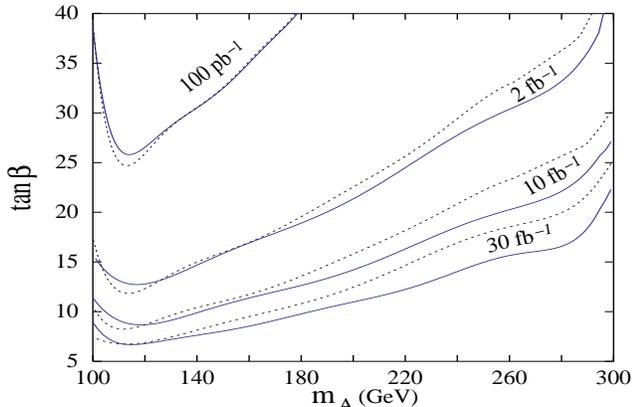}\\[-0.3cm]
\caption[fig:fig4]
        {95\%~C.L.~exclusion~contours 
	within the MSSM.
        Solid (dashed) lines correspond to the case
        $m_{\tilde f} = |\mu| = M_2 = 1 \tev~(150 \gev)$. }
      \label{fig:matb}
\end{figure}
%

We may mention in passing that there do exist
constraints on the $m_A$--$\tan \beta$ plane from 
direct searches at both the LEP collider and 
the Fermilab Tevatron~\cite{DreGucRoy,DiaHeTaiYua,CarMreWag} other
than the 100 pb$^{-1}$ curve shown here.
However, most of these are outside the scale of the present figure. It is
obvious, therefore, that a $3b$ signal could be remarkably effective,
if not in finding a Higgs boson of the MSSM, at least in putting
stringent constraints on the parameter space.

To summarize, then, we have discussed Higgs resonances in final
states with three tagged $b$-jets at the Fermilab Tevatron. For
Yukawa couplings of a Higgs boson to $b\bar b$ pairs -- which are
enhanced with respect to the SM coupling, we predict large signals in
bins of invariant mass constructed using the hardest pair of jets.
These are used to define a model-independent constraint on the
parameter space, which is then translated to the case of a
specific model -- the MSSM.  We show that this signal
could lead to constraints on the MSSM parameter space that better
present constraints by a significant margin.

\bigskip

{\small DC would like to thank the HEP Division, Argonne National
Laboratory, for hospitality while this work was being carried
out. The work of AD is supported by the Council for Scientific
and Industrial Research, Govt. of India. 
SR acknowledges partial financial
support from the World Laboratory, Lausanne. }


\end{narrowtext}

\begin{references}
\bibitem{comment:branching} The exact ratios depend on the choice of
(current) quark masses, where some latitude is allowed by
experimental errors.
\bibitem{comment:gam_gam} The two-photon mode is not of interest here,
as it is suppressed even further in the event of enhanced Yukawa
couplings.
\bibitem{comment:tautau} Even though the possibility of a $\tau^+\tau^-$ 
channel is not considered here, it is not absolutely precluded; however, 
it is admittedly not an easy task.
\bibitem{comment:resolution} A binning of 10 GeV is not quite
consistent with the mass resolution discussed before, especially for
masses above 125 GeV.  Nevertheless, we choose it because it allows
us to read off the differential cross-section directly from
Fig.~\ref{fig:eventnum}.  A modified binwidth is then easily
accommodated.
\end{references}
\end{document}